\documentclass[aps,prd,preprint,groupedaddress,nofootinbib]{revtex4-1}

\usepackage{graphicx}
\usepackage{epsfig}
\usepackage{bm}
\usepackage{amsfonts}
\usepackage{amsmath}
\usepackage{amssymb}

\begin{document}

\title{Structure Formation in Dark Matter Particle Production Cosmology}

\author{
Z. Safari$^{1}$\footnote{z.safari@sci.uok.ac.ir},
K. Rezazadeh$^{2}$\footnote{kazem.rezazadeh@ipm.ir},
and B. Malekolkalami$^{1}$\footnote{b.malakolkalami@uok.ac.ir}
}

\affiliation{
$^{1}$\small{Department of Physics, University of Kurdistan, Pasdaran Street, P.O. Box 66177-15175, Sanandaj, Iran}\\
$^{2}$\small{School of Physics, Institute for Research in Fundamental Sciences (IPM),
P.O. Box 19395-5531, Tehran, Iran}\\
}

\date{\today}


\begin{abstract}

We investigate a cosmological scenario in which the dark matter particles can be created during the evolution of the Universe. By regarding the Universe as an open thermodynamic system and using non-equilibrium thermodynamics, we examine the mechanism of gravitational particle production. In this setup, we study the large-scale structure (LSS) formation of the Universe in the Newtonian regime of perturbations and derive the equations governing the evolution of the dark matter overdensities. Then, we implement the cosmological data from Planck 2018 CMB measurements, SNe Ia and BAO observations, as well as the Riess et al. (2019) local measurement for $H_0$ to provide some cosmological constraints for the parameters of our model. We see that the best case of our scenario ($\chi_{{\rm tot}}^{2}=3834.40$) fits the observational data better than the baseline $\Lambda$CDM model ($\chi_{{\rm tot}}^{2} = 3838.00$) at the background level. We moreover estimate the growth factor of linear perturbations and show that the best case of our model ($\chi_{f\sigma_{8}}^{2}=39.85$) fits the LSS data significantly better than the $\Lambda$CDM model ($\chi_{f\sigma_{8}}^{2}=45.29$). Consequently, our model also makes a better performance at the level of the linear perturbations compared to the standard cosmological model.


\end{abstract}

\maketitle


\section{Introduction}
\label{section:introduction}

One of the factors that play a key role in the description of the physics of the early Universe is non-equilibrium thermodynamics. Production of matter due to the space-time reactions leads to the growth of entropy, while the reverse process is not allowed theoretically. In order to understand the effects of the matter creation process on the evolution of the Universe, a great deal of effort has gone. The impact of particle production on the evolution of the expanding Universe was studied by Schr\"{o}dinger, for the first time \cite{Schrodinger}, by using the microscopic description of the gravitational production of particles. According to his suggestion, in consequence of the effects of the gravitational field on the quantum vacuum, the particles can be continually created as the Universe expands. Later on, to find new consequences of the quantum field theory for the fundamental particles, this idea was utilized again based on the quantum field theory considerations in the curved space-time, by Parker and others \cite{Parker, Birrell, Mukhanov, Grib}. They pointed out that an equal amount of matter and antimatter would be created through this mechanism. Also, they argued that although the formation of particles within the expanding Universe is currently trivial, it might be of great cosmic importance within the earlier stages of the Universe's evolution.

Until 1988, many authors used the macroscopic Zel'dovich view \cite{Zeldovich:1969sb} that the matter creation could be simulated by the bulk viscosity mechanism. However, such a representation does not make sense thoroughly and the basic differences between bulk viscosity and macroscopic matter creation were discussed in \cite{Lima:1992np}. In 1989, a cosmological model was proposed based on the study of large-scale entropy production by Prigogine et al. in which particles were continuously produced due to the expansion of the Universe \cite{Prigogine}. Considering the thermodynamics of open systems, he inferred that Einstein's field equations confirm that particle production is possible in his scenario. Particle production in open systems results from non-equilibrium thermodynamics whose implications are consistent with general relativity. The creation of particles arising from the expansion of the Universe and consequently the entropy production, at the macroscopic level, will be possible via the redefinition of the momentum-energy tensor in Einstein's equations. In this case, the modified energy-momentum tensor satisfies the energy conservation law ($ T^{\mu \nu}_{~~;\nu}=0 $), but the cosmic fluid pressure has been modified because of the created particles. Prigogine et al. \cite{Prigogine} looked for a source to entropy production, but adiabatic reversible Einstein's equations failed to provide it. They applied thermodynamics in the context of cosmology and demonstrated that in the open thermodynamics systems, the energy-momentum tensor is naturally modified in a way that includes the production of matter and entropy at a macroscopic level. It is also the energy density and pressure of the cosmic fluid that determine the geometry of the Universe. One prominent point of this progress is that the particles are created only in the irreversible process, which represents an irreversible transfer of energy from the gravitational field to the created matter. In \cite{Lima:2007kk, Calvao}, the authors revisited the phenomenological approach to the mechanism of matter creation in the cosmological context in a covariant formulation. They demonstrated that the results of Prigogine et al. \cite{Prigogine} are valid if the specific entropy (entropy per particle) is constant. However, Prigogine et al. \cite{Prigogine} argued that the specific entropy should be constant ($ \dot{s}=0 $) because of the energy conservation law. The theory of general relativity permits the creation of particles through thermodynamics in open systems, and the creation of matter can be viewed as a source of internal energy. On the other hand, Gunzig et al \cite{Gunzig} determined the thermodynamical conditions required in the early and late time Universe. Particle production rate is not unique and $ \Gamma $ has different forms in each the Universe evolutionary era. For instance, $ \Gamma  \propto H^2$ satisfies true vacuum for radiation dominated era while $\Gamma \propto H^{-1}$ is a suitable choice for late time, and a simple choice for a decelerating Universe is $ \Gamma  \propto H$. In addition, particle production by black holes and its compatibility with the laws of thermodynamics have been studied elaborately by Hawking \cite{Hawking}. The original creation of cold dark matter model emulating  $\Lambda$CDM cosmology was proposed nearly a decade ago \cite{Lima:2009ic}. More recently, a relativistic kinetic formulation for such a model was discussed in \cite{Lima:2014hda}. A model with creation of baryon and cold dark matter particles is also investigated in \cite{Lima:2015xpa}, where the authors discussed the evolution of perturbations in the relativistic framework.

One might say that Newtonian cosmology commenced with the papers by Milne and McCrea \cite{Milne, McCrea}. In this approach, the uniform pressure does not perform a dynamical role in the continuity, Euler, and Poisson equations and so its generalization to the models involving pressure appeared to be inevitable. Lima et al.  \cite{Lima:1996at} investigated the cosmological perturbations in the Newtonian Universe once again, and they however did not ignore the pressure of the background fluid in their work. After eliminating the annoying pressure gradient term, they concluded that the resulting equations for the growth of density contrast in the homogeneous background with pressure are consistent with relativistic field equations. In this context, Reis \cite{Reis2003} revisited the cosmological models involving the time-dependent equation of state and in the presence of the non-adiabatic perturbations. He concluded that in such a case, the modified continuity equation suggested by Lima et al. \cite{Lima:1996at} cannot guarantee the compatibility between Newtonian and relativistic theories, and this approach is restricted to some specific cases with the assumption of adiabatic pressure perturbation.

In this paper, we aim to study the linear regime of perturbations in a cosmological framework involving dark matter particle production. Specifically, we derive the basic equations governing the dark matter inhomogeneities and apply them for some cases which have interesting motivations from both the theoretical and phenomenological perspectives. In our investigation, we compare the results of our model with those of the concordance $\Lambda$CDM model at both the background and linear perturbations levels. At the background level, we study the mechanism of particle production by applying the non-equilibrium thermodynamics on the homogeneous and isotropic Universe. In order to check the observational compatibility of our scenario, we apply the observational results from the Planck 2018 measurements of CMB temperature and polarization \cite{Planck:2018vyg, Planck:2019nip, Planck:2018lbu}, the Pantheon Supernovae (SN) sample \cite{Pan-STARRS1:2017jku}, the BAO measurements \cite{BOSS:2016wmc, Ross:2014qpa, Beutler:2012px}, and the Riess et al. (2019) measurement of the Hubble constant \cite{Riess:2019cxk}. Using the MCMC technique, we provide some cosmological constraints for the parameters of our model.

We moreover check the compatibility of our cosmological particle production scenario in light of the experimental data at the level of linear perturbations. To do so, we implement the perturbation equations that we derived for the density contrasts of the matter components to estimate the growth factor in our scenario and compare our results with the observational data from the LSS measurements. In this way, we can also check the compatibility of our model versus the $\Lambda$CDM model in light of the LSS data.

This paper is structured as follows. In Sec \ref{section:background}, we introduce the formulation of our setup and review the key equations governing its background dynamics. In Sec. \ref{section:perturbations}, we study the cosmological perturbations theory in a Newtonian Universe in the presence of non-vanishing fluid pressure. In this section, we derive in detail the evolutionary equation of density contrast in a model including dark matter particle production. Subsequently, in Sec \ref{section:constraints}, we present the cosmological constraints for our model using the observational data from different resources and compare the compatibility of our model in front of the baseline $\Lambda$CDM scenario at the background level. In Sec. \ref{section:growth_factor}, we utilize the perturbed equations derived in Sec. \ref{section:perturbations} to evaluate the matter density contrast during the Universe evolution. In particular, we compute the growth factor in our setup and compare our findings with the LSS observations at the level of linear perturbations. Eventually, we summarize our concluding remarks in Section \ref{section:conclusions}.

 
\section{The setup}
\label{section:background}

To take into account the contribution of particle production during the evolution of the Universe, one can modify the energy-momentum tensor for a relativistic fluid as follows
\begin{equation}
T_{\mu \nu} = (\rho +P + \Pi) u_{\mu} u_{\nu} + (P + \Pi) g_{\mu \nu},
\label{emtensor}
\end{equation}
where $u_{\mu}$ denotes the four-vector of velocity satisfying $u_{\mu}u^{\mu}=-1$. In the present work, the homogeneous and isotropic background is described by the FRW metric so that $\Theta\equiv u_{~~;\mu}^{\mu}=3\dot{a}/a=3H$, and $\dot n = n,_{\mu} u^\mu $. Here, $a$ is used for the scale factor of the Universe, $H$ represents the Hubble expansion rate, and the dot indicates the derivative with respect to the cosmic time. Note that $\rho$ and $P$ are the energy density and equilibrium pressure of the content of the Universe, respectively. The contribution of the particle production is exerted by the creation pressure $\Pi$. In this regard, one should apply two conservation laws, namely the conservation of particle number ($(N^{\mu} = n u^{\mu})_{;\mu} = 0$) in a closed thermodynamics system, and the conservation of energy ($T^{\mu \nu}_{~~~; \nu} = 0$). These conservation rules lead to the following equations in our scenario
\begin{align}
&\dot n + \Theta n = 0, \label{ndot} \\ 
&\dot \rho + \Theta (\rho +P + \Pi)=0. \label{rhodot}
\end{align}
But if the Universe is considered as an open thermodynamic system, the particle number will no longer remain constant. As a result, Eq. \eqref{ndot} should be modified as follows
\begin{equation}
\dot n + \Theta n = n \Gamma,
\label{ndot2}
\end{equation}
where $\Gamma$ is the particle production rate whose explicit expression is determined by the quantum field theory. Modifying the conservation law of particle number, Gibbs' equation is further modified as
\begin{equation}
\dot \rho + \Theta \left( 1 - \frac{\Gamma}{\Theta} \right)(\rho + P) = n T \dot s,
\end{equation}
where $T$ is the fluid temprature and $ s=S/N$ is the specific entropy (entropy per particle). Under the adiabatic condition $  \dot{s}=0 $ the standard continuity equation can  be recovered when $ \Gamma \ll \Theta $. In the special case under which $ \Gamma = \Theta$, regardless of the amount of equation of state, the energy density will be constant $ \dot{\rho} =0 $, and the de-Sitter phase will occur. Also in such a case $ \dot{n}=0 $ and thermodynamic equilibrium will be established. Assuming that the process of particle creation occurs adiabatically ($\dot s=0$), the creation pressure is thus obtained in the following form
\begin{equation}
\Pi = - \frac{\Gamma}{\Theta} (\rho + P).
\label{pi}
\end{equation}
Using the modified momentum-energy tensor, the field equations in the flat FRW  metric with $ \Theta = 3 \dot{a}/a $ turn into
\begin{align}
\label{Friedmann}
& \frac{\dot{a}^2}{a^2} = \frac{8 \pi G }{3}  \rho, \\ 
\label{acceleration}
& \frac{\ddot{a}}{a}=-\frac{4\pi G}{3}\left[\rho+3(P+\Pi)\right],\\
\label{continuity}
& \dot{\rho_i}+ 3 \frac{\dot{a}}{a} (\rho + P + \Pi) = 0. 
\end{align}
In the case of the cold dark matter component ($P_c = 0$), with the following creation pressure,
\begin{equation}
\Pi_c = - \frac{\Gamma}{\Theta} \rho_c,
\end{equation}
the continuity equation and the evolution equation of the dark matter density are resulted in as
\begin{align}
&\dot{\rho}_{c}=- \Theta \rho_{c}\left[1-\frac{\Gamma}{\Theta}\right],  \\
&\rho_{c}=\rho_{c0}a^{-3}\exp\left[3\int_{1}^{a}\frac{\Gamma}{\Theta}\frac{da}{a}\right].
\label{matter}
\end{align}
Finally, in the cosmological setting including dark matter particle production, the expansion of the Universe is described by the following equation
\begin{align}
\frac{H^{2}}{H_{0}^{2}}=\Omega_{b}~a^{-3}+\Omega_{c}~a^{-3}\exp\left[3\int_{1}^{a}\frac{\Gamma}{\Theta}\frac{da}{a}\right]+\Omega_{\Lambda},
\label{ourmodel}
\end{align}
where $\Omega_{b}$, $\Omega_{c}$, and $\Omega_{\Lambda}$ denote the density parameters for baryonic matter, cold dark matter, and cosmological constant, respectively. The background expansion rate is denoted with  $H$ ($\equiv \dot{a}/a$) and the local expansion rate is denoted with $\Theta$ ($\equiv u^{\mu}_{~; \mu}$). In this work we use flat FRW metric, thus $ \Theta = 3 \dot{a}/a$.

To go ahead, it is necessary to know the explicit form of the particle production rate. The main approach to determine the particle production rate is to apply the quantum field theory implications in curved space-time. Since the nature of the produced particles affects the production rate, and in addition, the nature of dark matter is still unknown for us, some researchers apply the phenomenological forms for $\Gamma$ \cite{Gunzig, Shapiro, Nunes}. From Eqs. \eqref{pi} and \eqref{Friedmann}, it can be easily inferred that $\Gamma = 3H$ leads to the de-Sitter late time ($\dot{\rho}=0, \dot{H} = 0$), regardless of the equation of state for the matter-energy content of the Universe. A general phenomenological choice for particle production rate during the accelerated phase is $\Gamma \propto H$. Following \cite{Nunes}, we consider the three following functional expressions for the production rate in our investigation:
\begin{align}
& \textrm{Model 1:} \qquad \Gamma= 3 \beta H,
\label{Model1}
\\
& \textrm{Model 2:} \qquad \Gamma= 3 \beta H  \left[5-5\tanh(10-12a)\right],
\label{Model2}
\\
& \textrm{Model 3:} \qquad \Gamma= 3 \beta H  \left[5-5\tanh(12a-10)\right],
\label{Model3}
\end{align}
where $\beta$ is a positive constant. In the following sections, we examine these models at the background and perturbations levels and compare their implications in light of the recent observational data.


\section{Cosmological perturbations}
\label{section:perturbations}

The neo-Newtonian approach which is suggested by McCrea \cite{McCrea1951}, is based on the following equations
\begin{align}
&\frac{ \partial  \rho_i}{\partial t} + \vec{\nabla}_r . \left[\left( \rho_i + P_i\right) \vec{u_i} \right] = 0, \label{continuity1}  \\
& \frac{ \partial  \vec{u_i}}{\partial t} +  \vec{u_i}. \vec{\nabla}_r \vec{u_i}= - \vec{\nabla}_r \Psi - \frac{\vec{\nabla}_r P_i}{\rho_i + P_i}, \label{Euler1} \\
&\nabla_{r}^{2}\Psi=4\pi G\sum_{i}\left(\rho_{i}+3P_{i}\right). \label{Poisson1}
\end{align}
where $\rho$, $P$, $\vec{u}$, and $\Psi$ are the energy density, pressure, field velocity, and generalized gravitational potential of the perfect fluid, respectively. Within a homogeneous and isotropic Universe ($P=P(t), \rho = \rho(t)$), the velocity of the fluid is given by Hubble's law
\begin{equation}
\vec{u} = \frac{\dot{a}}{a} \vec{r},
\end{equation}
where $\vec{r}$ is the physical distance. In such a case, the Friedmann equations derived from Einstein's gravitational field equations, describe the evolution of the scale factor as follows
\begin{align}
\frac{\dot{a}^2}{a^2} & = \frac{8 \pi G }{3}\sum_i \rho_i, \\
\frac{\ddot{a}}{a} & = - \frac{4 \pi G}{3} \sum_i (\rho_i +3P_i).  
\end{align}
These equations are valid for the sum of components of the Universe. However, the continuity equation for this homogeneous and isotropic Universe takes the following form
\begin{equation}
\frac{\partial\rho_{i}}{\partial t}+3\frac{\dot{a}}{a}\left(\rho_{i}+P_{i}\right)=0.
\end{equation}
This equation is valid for each component of the Universe, separately.

Using these equations to study the perturbed space-time leads to disagreement with the corresponding equations in the relativistic approach. The density contrast equation obtained in this manner is not consistent with the corresponding relativistic equation in the synchronous gauge. Lima et al. \cite{Lima:1996at} argued that the root of this problem lies in the continuity equation. Following Peebles \cite{Peebles1993}, they changed the partial time derivative at a fixed physical distance ($\vec{r}$) to a partial time derivative at a fixed comoving distance ($\vec{x}$). The two partial time derivatives are related together as follows
\begin{align}
\vec{r}(t) &= a(t) \vec{x}, \\ 
\vec{u}_0&= \dot{a} \vec{x}, \\
\vec{\nabla}_r &= \frac{1}{a} \vec{\nabla}_x, \\
\left( \frac{\partial}{\partial t} \right)_r &= \left( \frac{\partial}{\partial t} \right)_x - \frac{\dot{a}}{a} \left( \vec{x}. \vec{\nabla}_x \right). \label{comoving}
\end{align}
According to the usual procedure in the cosmological perturbations theory, let us consider the small fluctuations ($ \delta\rho, \delta P, \phi, \vec{v} $) around the homogeneous background quantities ($ \rho_0,  P_0, \Psi_0, \vec{u}_0 $) as follows
\begin{align}
 \rho_{i} &= \rho_{0_{i}}(t)\left[1+\delta_{i}(\vec{r},t)\right],  
 \label{rho-perturb} \\
 P_{i} &= P_{0_{i}}(t)+\delta P_{i}(\vec{r},t),
 \label{P-perturb} \\
  \vec{u_{i}} &= \vec{u}_{0_{i}}+\vec{v}_{i}(\vec{r},t),
 \label{u-perturb}\\
  \Psi &= \Psi_0 (\vec{r},t) + \phi (\vec{r},t), 
  \label{Psi-perturb} 
\end{align}
where the zero index represents the background quantities, and $ \delta_j=\delta\rho_j/\rho_{0j} $ denotes the density contrast. By inserting Eqs. \eqref{rho-perturb}-\eqref{u-perturb} into Eqs. \eqref{continuity1}, \eqref{Euler1}, and \eqref{Poisson1}, and to the first order of perturbations, we arrive at the following equations
\begin{align}
& \rho_{0_{i}}\left[\left(\frac{\partial\delta_{i}}{\partial t}\right)_{r}+\vec{u_{0_{i}}}.\vec{\nabla}_{r}\delta_{i}\right]-3\frac{\dot{a}}{a}P_{0_{i}}\delta_{i}+3\frac{\dot{a}}{a}\delta P_{i}+\left(\rho_{0_{i}}+P_{0_{i}}\right)\vec{\nabla}_{r}.\vec{v_{i}}=0, 
\label{Eqs1} \\
& \left(\frac{\partial\vec{v_{i}}}{\partial t}\right)_{r}+\left(\vec{u_{0_{i}}}.\vec{\nabla}_{r}\right)\vec{v_{i}}+\vec{v_{i}}.\vec{\nabla}_{r}\vec{u_{0_{i}}}=-\vec{\nabla}_{r}\phi-\frac{\vec{\nabla}_{r}\delta P_{i}}{\rho_{0_{i}}+P_{0_{i}}}, 
\label{Eqs2} \\
&\nabla_{r}^{2}\phi=4\pi G\sum_{i}\left(\delta\rho_{i}+3\delta P_{i}\right). \label{Eqs3}
\end{align}
The equation of state parameter for a perfect fluid and the speed of sound parameter are always defined in terms of the background quantities ($\rho_0$, $P_0 $), but the effective speed of sound is defined in terms of the perturbed quantities $\delta\rho$ and $\delta P$. In the following equations, one can see how these quantities are related to each other,
\begin{align}
c_{\mathrm{eff}_{i}}^{2} &= \frac{\delta P_{i}}{\delta\rho_{i}}, \\
\frac{\delta P_{i}}{\rho_{0_{i}}} &= \frac{c_{\mathrm{eff}_{i}}^{2}\delta\rho_{i}}{\rho_{0_{i}}} =c_{\mathrm{eff}_{i}}^{2}\delta_{i},  \\
c_{s_{i}}^{2} &= \frac{\dot{P}_{0_{i}}}{\dot{\rho}_{0_{i}}}, \\
\omega_{i} &= \frac{P_{0_{i}}}{\rho_{0_{i}}},  \\
\dot{\omega}_{i} &= -3H\left(1+\omega_{i}\right)\left(c_{s_{i}}^{2}-\omega_{i}\right).
\label{ceff}
\end{align}
Considering particle production, we can re-interpret the effective sound speed as follows
\begin{equation}
c_{\mathrm{eff}}^{2}=\frac{\delta (P + \Pi)}{\delta \rho}.
\end{equation}
In Eq. \eqref{pi}, if $\Gamma$ is linearly related to $ H $, then the perturbations in creation pressure does not depend on $ \delta H $, but they depend only on $ \delta \rho $ and $\delta P$. Therefore, the first order perturbations do not include $ \delta H $ as long as $\Gamma$ is linearly related to $ H $.

With the help of Eqs. \eqref{comoving} and the above equations, it is easily demonstrated that Eqs. \eqref{Eqs1}-\eqref{Eqs3} are obtained as follows
\begin{align}
&\dot{\delta_{i}}+3\frac{\dot{a}}{a}\left(c_{\mathrm{eff}_{i}}^{2}-\omega_{i}\right)+\frac{\left(1+\omega_{i}\right)}{a}\vec{\nabla}.\vec{v_{i}}=0,
\label{Eq.delta}\\
&\dot{\vec{v}}_{i}+\frac{\dot{a}}{a}\vec{v}_{i}=-\frac{1}{a}\vec{\nabla}\phi-\frac{1}{a}\frac{c_{\mathrm{eff}_{i}}^{2}}{\left(1+\omega_{i}\right)}\vec{\nabla}\delta_{i}, 
\label{Eq.euler} \\
&\nabla^{2}\phi=4\pi Ga^{2}\sum_{i}\rho_{0_{i}}\delta_{i}\left(1+3c_{\mathrm{eff}_{i}}^{2}\right). 
\label{Eq.poison}
\end{align}
These equations are in agreement with the equations that are used to study the large-structure formation in the Newtonian regime of perturbations (see, e.g., \cite{Lima:1996at, Hwang:1997xt, Hwang:2005xt, Abramo:2008ip, Rezaei:2017yyj, Fahimi:2018pcr, Rezazadeh:2020zrd}).

Here, it is useful to replace $\left( \frac{\partial \delta}{\partial t} \right)_x = \dot{\delta}$, $\left( \frac{ \partial  \vec{v}}{\partial t}\right) _x = \dot{ \vec{v}}$, $\vec{\nabla}_x = \vec{\nabla}$. Eliminating the peculiar velocity from Eqs. \eqref{Eq.delta} and \eqref{Eq.euler}, and also substituting Eq. \eqref{Eq.poison}, we acquire the following differential equation describing the evolution of density contrast
\begin{align}
& \ddot{\delta}_{i}+\dot{\delta}_{i}\left[H\left(3c_{\mathrm{eff}_{i}}^{2}-3\omega_{i}+2\right)-\frac{\dot{\omega}_{i}}{\omega_{i}+1}\right]
\nonumber \\
& +3H\delta_{i}\left[2c_{\mathrm{eff}_{i}}\dot{c}_{\mathrm{eff}_{i}}-c_{\mathrm{eff}_{i}}^{2}\left(2H+\frac{\dot{H}}{H}-\frac{\dot{\omega}_{i}}{\omega_{i}+1}\right)-2H\omega_{i}+\omega_{i}\frac{\dot{H}}{H}-\frac{\dot{\omega}_{i}}{\omega_{i}+1}\right]
\nonumber \\
& +\frac{k^{2}c_{\mathrm{eff}_{i}}^{2}\delta_{i}}{a^{2}}-\frac{3}{2}H^{2}\left(\omega_{i}+1\right)\sum_{j}\left(1+3c_{\mathrm{eff}_{j}}^{2}\right)\Omega_{j}\delta_{j}=0.
\label{ddotdeltai}
\end{align}
In our study for the matter perturbations, we consider the baryonic matter and cold dark matter perturbations separately. It should be noted that due to the particle production, the equation of state of the cold dark matter is not constant here. In addition, since we assume the adiabatic perturbations, therefore we can approximate the effective sound speed of the cold dark matter as its adiabatic sound speed, $c_{\mathrm{eff}_{c}}^{2}\approx c_{s_{c}}^{2}=\omega_{c}-\dot{\omega}_{c}/3H\left(\omega_{c}+1\right)$, which is valid in the linear regime of perturbations up to a good approximation. With these assumptions, the equations for the baryonic matter and cold dark matter perturbations are obtained from Eq. \eqref{ddotdeltai} respectively as
\begin{align}
& \ddot{\delta}_{b}+2H\dot{\delta}_{b}-\frac{3}{2}H^{2}\left[\Omega_{b}\delta_{b}+\left(1+3c_{s_{c}}^{2}\right)\Omega_{c}\delta_{c}\right]=0
\label{ddotdeltab} \\
& \ddot{\delta}_{c}+\dot{\delta}_{c}\left[H\left(3c_{s_{c}}^{2}-3\omega_{c}+2\right)-\frac{\dot{\omega}_{c}}{\omega_{c}+1}\right]
\nonumber \\
& +3H\delta_{c}\left[2c_{s_{c}}\dot{c}_{s_{c}}-c_{s_{c}}^{2}\left(2H+\frac{\dot{H}}{H}-\frac{\dot{\omega}_{c}}{\omega_{c}+1}\right)-2H\omega_{c}+\omega_{c}\frac{\dot{H}}{H}-\frac{\dot{\omega}_{c}}{\omega_{c}+1}\right]
\nonumber \\
& +\frac{k^{2}c_{s_{c}}^{2}\delta_{c}}{a^{2}}-\frac{3}{2}H^{2}\left(\omega_{c}+1\right)\left[\Omega_{b}\delta_{b}+\left(1+3c_{s_{c}}^{2}\right)\Omega_{c}\delta_{c}\right]=0.
\label{ddotdeltac}
\end{align}
These equations are the two coupled equations that we should solve simultaneously to determine the evolutions of the baryonic matter and cold dark matter density contrasts which are denoted by $\delta_b$ and $\delta_c$, respectively. Then, we use the solutions for $\delta_b$ and $\delta_c$ in the following equation to calculate the matter density contrast
\begin{equation}
 \label{deltam}
 \delta_{m}=\frac{\rho_{b}\delta_{b}+\rho_{c}\delta_{c}}{\rho_{b}+\rho_{c}}.
\end{equation}
We will use these equations in Sec. \ref{section:growth_factor} to estimate the growth factor in our scenario and compare our results with the observational data.


\section{Cosmological Constraints}
\label{section:constraints}

In this section, we are interested in constrain our model observationally at the level of background dynamics. For this purpose, we implement the CosmoMC package \cite{Lewis2000, Lewis2002} to estimate the seven free parameters of the model (\ref{ourmodel}), including $\{ \Omega_b h^2, \Omega_c h^2, \theta_{MC}, \tau, A_s, n_s, \beta \}$, where $\Omega_b$ is the present baryon density parameter, $\Omega_c$ is the present dark matter density parameter, $\theta_{MC}$ is the approximation to the ratio of comoving size to comoving angular diameter distance, $\tau$ is the optical depth, $A_s$ is the amplitude of the scalar power spectrum, $n_s$ is the scalar spectral index, and $\beta$ is the production rate parameter. We suppose flat priors on these parameters in our numerical analysis. Following the Planck collaboration, we suppose free-streaming neutrinos as two massless species and one massive with $M_{\nu}=0.06\,\mathrm{eV}$ \cite{SimonsObservatory:2018koc}.

The CosmoMC package \cite{Lewis2000, Lewis2002} uses Markov Chain Monte Carlo (MCMC) algorithm to calculate the likelihood of cosmological parameters by using the observational data from different resources. Multiplying the separate likelihoods of CMB, SNe Ia, BAO, and Riess et al. (2019) data gives us the total likelihood $\mathcal{L}\propto e^{-\chi_{{\rm tot}}^{2}/2}$, where $\chi_{{\rm tot}}^{2}=\chi_{\mathrm{CMB}}^{2}+\chi_{\mathrm{SN}}^{2}+\chi_{\mathrm{BAO}}^{2}+\chi_{\mathrm{Riess2019}}^{2}$ represents the difference between observational value and theoretical value (for more details about cosmological constraints see \cite{Karami2013, Karami2014}). Following \cite{Poulin:2018cxd, Smith:2019ihp, Poulin:2021bjr, Murgia:2020ryi}, we put the upper bound on the Gelman-Rubin convergence criterion \cite{Gelman:1992zz} as $R-1<0.1$ in our MCMC analysis.

For the CMB data in our MCMC analysis, we include the Planck 2018 \cite{Planck:2018vyg, Planck:2019nip, Planck:2018lbu} measurements for the anisotropies in temperature and polarization spectra of the CMB radiation. The acoustic peaks of the temperature power spectrum of the cosmic microwave background radiation provide useful information about the expansion history of the Universe. The physics of decoupling affects the amplitude of the acoustic peaks and the physics of between the present and the decoupling epoch changes the locations of peaks. We use the Planck 2018 measurements of CMB temperature and polarization at small (TT,TE,EE) and large angular scales (lowl+lowE) \cite{Planck:2018vyg, Planck:2019nip}. We also include the Planck CMB lensing potential power spectrum in the multipole range $40\leq\ell\leq400$ \cite{Planck:2018lbu}.

Since type Ia supernovae have the same absolute magnitude, these standard candles are a powerful tool for exploring the history of the expansion of the Universe. In our MCMC analysis, we employ the Pantheon SN sample \cite{Pan-STARRS1:2017jku}, which is comprised of measurements of the luminosity distances of 1048 SNe Ia in the redshift interval $0.01 < z < 2.3$.

Another powerful tool to probe the expansion history of the Universe is the BAO's standard ruler. The anisotropies in CMB and large-scale structures of matter are affected by the pressure waves coming from the cosmological perturbations in baryon-photon primordial plasma. The observed peak in the large-scale correlation function measured by the luminous red galaxies of Solon Digital Sky Survey (SDSS) at $z=0.35$ \cite{Ross:2014qpa} and $z=0.278$ \cite{Kazin} reveals the baryon acoustic oscillations at $100h^{-1}\,\mathrm{Mpc}$ as well as in the two-degree Field Galaxy Survey (2dFGS) at $z=0.2$ \cite{Percival}, six-degree Field Galaxy Survey (6dFGS) at $z=0.106$ \cite{Beutler:2012px}, $z=0.44, z=0.60$ and $z=0.73$ by WiggleZ team \cite{Blake}, the SDSS Data Releases 7 main Galaxy sample at $z=0.15$ \cite{Ross}, the Data Releases 10 and 11 Galaxy samples at $z=0.57$ \cite{Anderson}. In our work, we consider the BAO dataset from BOSS DR12 \cite{BOSS:2016wmc}, SDSS Main Galaxy Sample \cite{Ross:2014qpa}, and 6dFGS \cite{Beutler:2012px}.

Another independent constraint that can be applied to the estimation of the model parameters is the local measurements for the present Hubble parameter. In the present work, we include the Riess et al. (2019) constraint on the Hubble constant, $H_0 = 74.03 \pm 1.42\,{\rm km\,s^{-1}\,Mpc^{-1}}$ \cite{Riess:2019cxk}, which is provided by the Hubble Space Telescope (HST) observations of $70$ long-period Cepheids in the Large Magellanic Cloud.

Using the computational package of CosmoMC \cite{Lewis2000, Lewis2002}, we explore the parameter space for the three models introduced in Sec. \ref{section:background}, and generate a set of MCMC chains. To analyze the MCMC chains, we use the GetDist package \cite{Lewis:2019xzd} which is publicly available.

We perform a joint analysis including the datasets explained above, and obtain the confidence intervals and the best-fit values of the free parameters for the three models including dark matter particle production and $\Lambda$CDM without particle production ($ \beta = 0 $). The best-fit values and also the 68\% confidence level (CL) constraints for the parameters of the studied models have been summarized in Table \ref{table:parameters}. In the table, we also preset the values of some of the derived parameters including $H_0$, $\Omega_m$, $\Omega_\Lambda$, $\sigma_8$, and $S_8$.

The minimum values of $\chi^2$ for the models and the considered datasets are presented in Table \ref{table:chi2}. From the table, we infer that the minimum value of $\chi_{{\rm tot}}^{2}$ belongs to Model 2, and therefore this model provides the best fit with the CMB, SN, BAO, and Riess et al. (2019) data in comparison with the other models. In particular, the value of $\chi_{{\rm tot}}^{2}$, in this case, is reduced considerably relative to the $\Lambda$CDM scenario, and this point implies that the particle production scenario fits the recent observational data better the standard cosmological scenario. The better performance for Model 2 in fitting to the observational data may originate somewhat from the additional degree of freedom which is the production rate parameter $\beta$, but an improvement of $\Delta \chi^2 = -3.6$ compared to $\Lambda$CDM with only one additional degree of freedom is somewhat interesting and worth further investigation. Model 3 fits the data better than $\Lambda$CDM but its improvement is not as significant as the one for Model 2. Model 1, however, fails to fit the data better than $\Lambda$CDM.

\begin{table}[!ht]
 
\caption{The best-fit values and 68\% CL constraints for the parameters of the investigated models.}
\centering

 \scalebox{0.7}{
\begin{tabular} {|c|l  c|l  c|l  c|l  c|}
\hline
Parameter & \multicolumn{2}{c|}{\begin{tabular}{c c}
\multicolumn{2}{c}{Model 1} \\ 
best-fit~~~~~~~~~~~~~ & 68\% limits \\ 
\end{tabular}} & \multicolumn{2}{c|}{\begin{tabular}{c c}
\multicolumn{2}{c}{Model 2} \\ 
best-fit~~~~~~~~ & 68\% limits \\ 
\end{tabular}} & \multicolumn{2}{c|}{\begin{tabular}{c c}
\multicolumn{2}{c}{Model 3} \\ 
best-fit~~~~~~~~~~~~~ & 68\% limits \\ 
\end{tabular}} & \multicolumn{2}{c|}{\begin{tabular}{c c}
\multicolumn{2}{c}{$\Lambda$CDM} \\ 
best-fit~~~~~~~~ & 68\% limits  \\ 
\end{tabular}} \\ 
\hline 

$\Omega_b h^2   $ & $ 0.022457 $ & $0.02250\pm 0.00015$ &  $ 0.0224034 $& $0.02243^{+0.00012}_{-0.00020}$ & $ 0.0224437 $ & $0.02250\pm 0.00014 $ & $0.02258$& $0.02252\pm 0.00013$\\

$\Omega_c h^2   $ & $0.117666$ & $0.1177\pm 0.0010          $ & $0.117788$ &$0.11892^{+0.00088}_{-0.0011}$ & $0.117902$ &$0.1177^{+0.0011}_{-0.00089}$& $0.118773$ &$0.11822\pm 0.00088$\\

$100\theta_{MC} $ & $1.04155$ &$1.04139^{+0.00032}_{-0.00038}$ & $1.04114$ & $1.04130\pm 0.00036$ & $1.04092$ &$1.04139^{+0.00029}_{-0.00040}$ & $1.0413$ & $1.04118\pm 0.00029$\\

$\tau           $ & $0.0617843$ & $0.0605\pm 0.0080          $ & $0.0579543$ & $0.0573^{+0.0068}_{-0.0079}$ & $0.0604604$ & $0.0599^{+0.0069}_{-0.0078}$ & $0.0550124$ & $0.0594\pm 0.0074$ \\

${\rm{ln}}(10^{10} A_s)$ & $3.05348$ & $3.052\pm0.016$ & $3.05058$ & $3.050_{-0.015}^{+0.014}$ & $3.05528$ & $3.052_{-0.015}^{+0.014}$ & $3.04439$ & $3.051\pm0.014$\\

$n_s$ & $0.9694$ & $0.9699\pm0.0039$ & $0.969395$ & $0.9697_{-0.0040}^{+0.0036}$ & $0.97236$ & $0.9699\pm0.0036$ & $0.969018$ & $0.9694\pm0.0037$\\

$\beta$ &$ 5.66354\times 10^{-5} $ & $< 0.000184                $ & $0.00175062$ & $ 0.0031^{+0.0013}_{-0.0017} $ &$ 8.29533\times 10^{-6} $ & $< 0.0000183               $ & $ - $ & $ - $\\

\hline

$H_0                       $ & $68.3063$ & $67.96^{+0.50}_{-0.40}     $ & $68.7384$ & $68.79\pm 0.59            $ & $67.9181$ & $67.93^{+0.53}_{-0.41}     $ & $68.0866$ & $68.20^{+0.42}_{-0.38}     $\\

$\Omega_\Lambda            $ & $0.698295$ & $0.6949^{+0.0058}_{-0.0048}$ & $0.701931$ & $0.6999\pm 0.0067     $ & $0.694354$ & $0.6947\pm 0.0057$ & $0.693692$ & $0.6959^{+0.0055}_{-0.0049}$\\

$\Omega_m$ & $0.301705$ & $0.3051^{+0.0048}_{-0.0058}$ & $0.298069$ & $0.3001\pm 0.0067           $ & $0.305646$ & $0.3053\pm 0.0057$ & $0.306308$ & $0.3041^{+0.0049}_{-0.0055}$\\

$\sigma_8 $ & $ 0.808233 $& $0.8073^{+0.0058}_{-0.0066}$& $ 0.813362 $& $0.8217^{+0.0081}_{-0.0096}$& $ 0.810012 $& $0.8070\pm 0.0064 $& $ 0.807798 $& $0.8090\pm 0.0058  $\\

$S_8  $ & $ 0.810526 $& $0.8140^{+0.0091}_{-0.011} $& $ 0.81074 $& $0.822\pm 0.013  $& $ 0.817598 $& $0.814\pm 0.010 $& $ 0.816248 $& $0.8144\pm 0.0098 $\\

\hline
\end{tabular}
}
\label{table:parameters}
\end{table}

\begin{table}[!ht]
 
\caption{The minimum value of $\chi^2$ for each model and each dataset. The values of $\chi^2_{\rm tot} $ and $\Delta \chi^2  = \chi^2_{\rm Model} -\chi^2_{\rm \Lambda CDM}$ are also presented in the table.}
\centering

 \scalebox{0.7}{
\begin{tabular} {|c|l  c|l  c|l  c|l  c|}
\hline
Parameter & \multicolumn{2}{c|}{\begin{tabular}{c c}
\multicolumn{2}{c}{Model 1} \\ 
best-fit~~~~~& 68\% limits \\ 
\end{tabular}} & \multicolumn{2}{c|}{\begin{tabular}{c c}
\multicolumn{2}{c}{Model 2} \\ 
best-fit~~~~~& 68\% limits \\ 
\end{tabular}} & \multicolumn{2}{c|}{\begin{tabular}{c c}
\multicolumn{2}{c}{Model 3} \\ 
best-fit~~~~~& 68\% limits \\ 
\end{tabular}} & \multicolumn{2}{c|}{\begin{tabular}{c c}
\multicolumn{2}{c}{$\Lambda$CDM} \\ 
best-fit~~~~~& 68\% limits  \\ 
\end{tabular}} \\ 
\hline 

$\chi^2_{\rm CMB}          $ & 2782.67 &$2796\pm 61                $ &2779.45 & $2815\pm 220                $ & 2778.76 & $2793\pm 31                 $&2780.39 & $2791\pm 12                $\\

$\chi^2_{\rm SN}          $ &1034.75 & $1034.91\pm 0.79           $ &1034.78 & $1034.94\pm 0.42            $ &1034.83 &$1034.88\pm 0.20           $&1034.85 & $1034.84\pm 0.14           $\\

$\chi^2_{\rm BAO}          $ &5.32163 & $5.7\pm 3.1                $ &6.28641 & $7.0\pm 3.1                 $ & 5.21791 &$5.56\pm 0.61               $&5.24274 & $5.59\pm 0.52              $\\

$\chi^2_{\rm Riess2019}$ &16.2473 & $18.4\pm 4.9$ &13.8866 & $13.8\pm 2.8$ &18.5258 & $18.6\pm 3.4          $&17.5187& $16.9\pm 2.3$\\

$\chi^2_{\rm tot}          $ &3838.98893 & $ -  $ & 3834.40301 & $ - $ &  3837.33371  &$- $& $3838.00144$ & $ -  $\\

$\Delta \chi^2          $ &0.98749 & $ - $ & -3.59843 & $ - $ & -0.66773 &$ - $& $0.0$ & $-$\\

\hline
\end{tabular}
}
\label{table:chi2}
\end{table}

\begin{figure}[t]
\begin{center}
\scalebox{0.6}[0.6]{\includegraphics{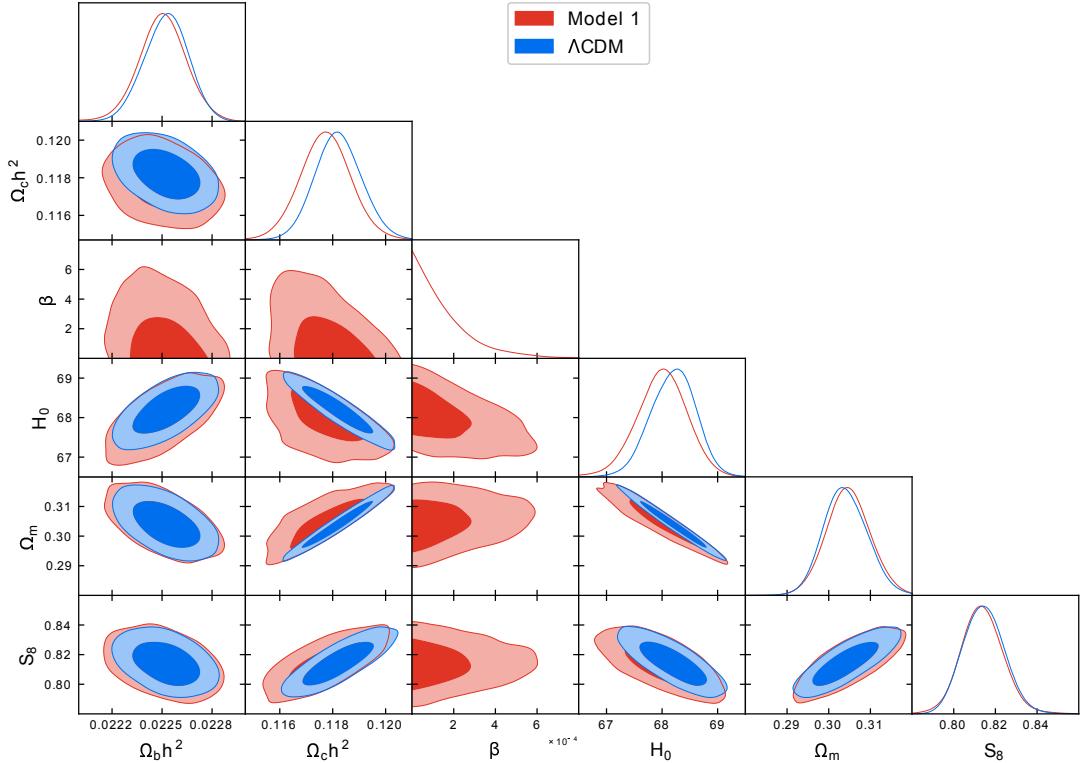}}
\caption{1D likelihoods and 2D contours for the parameters in 68\% and 95\% CLs for Model 1 (red) next to the $\Lambda$CDM constraints (blue).}
\label{figure:contour1}
\end{center}
\end{figure}

The 1D marginalized relative likelihood functions and 2D contours in 68\% and 95\% confidence intervals for Model 1 are shown in Figure \ref{figure:contour1}. As we see, the joint analysis puts a strong constraint on all of the parameters. On one hand, $\beta<0$ is not physically acceptable, because negative $\Gamma$ corresponds to $ \dot{S}<0 $, and violates the second law of thermodynamics. On the other hand, the best-fit value of $\beta$ is obtained within the 68\% CL region, and not at the beginning of the interval, so the estimated best-fit value of $\beta$ is reliable. The best-fit of the $\beta$ and its mean get very small values. However, although $ \beta $ takes very small values in Model 1, its non-vanishing value confirms the compatibility of the theory of particle creation with the recent observations.

In Table \ref{table:parameters}, we see that the best-fit and mean value of $S_ 8$ in Model 1 are slightly smaller than the corresponding values in the $\Lambda$CDM framework, and it seems that this model can reduce the $S_8$ tension a little in comparison with the $\Lambda$CDM scenario. However, to state this point we should be somewhat cautious because the $S_8$ parameter is an extrapolated quantity which results from the CMB and the cosmic shear data for the cosmic density fluctuations at very different redshifts. To extrapolate these data, it has been supposed that the number of matter particles remains conserved during the Universe evolution. Moreover, at the low redshifts, the cosmological perturbations enter in the nonlinear regime where $\delta_{m}\gtrsim1$, and therefore $\delta\rho_{m}$ becomes comparable with the background matter energy density $\bar{\rho}_{m}$. Therefore, the contribution of $\delta\rho_{m}$ should be included in the total matter energy density, and accordingly, today's matter density parameter $\Omega_m$ will be modified. In addition, from the definition $S_{8}=\sigma_{8}\sqrt{\Omega_{m}/0.3}$, the $S_8$ parameter is related to $\Omega_m$ directly, and consequently it will depend on the value of $\delta_m$. The dependency of $S_8$ to $\delta_m$ becomes more sensitive in the nonlinear regime of perturbations. Also, in addition to geometry, cosmic shear signals are sensitive to the growth of structures, and it is necessary to examine the nonlinear evolution of the Universe with greater precision to interpret the cosmic shear survey and consequently the $S_8$ tension. From these remarks, we conclude that the full estimation of the $S_8$ parameter requires more subtle treatment which is beyond the scope of the present work and is left for future investigations.
 
However, although in Model 1, we have assumed that the dark matter particle production occurs continuously during the Universe evolution, but as we see in Table \ref{table:parameters}, the parameter $\beta$ takes a very small value in this model. Hence the effects of particle production are negligible in this case.

In this scenario the best-fit and mean values of $ \Omega_b h^2 $ and $ \Omega_c h^2 $ are smaller than the $\Lambda$CDM results. The density parameter of the matter is the sum of the contributions for the cold dark matter and baryonic matter density parameters. The best-fit value of the  $ \Omega_m$ in this model is smaller than $ \Lambda$CDM, and consequently, the best-fit value of the $ \Omega_\Lambda$ in this model is bigger than the $\Lambda $CDM prediction. So, in Model 1, the contribution of dark energy in the Universe content is further than the one in $ \Lambda $CDM. 

\begin{figure}[t]
\begin{center}
\scalebox{0.6}[0.6]{\includegraphics{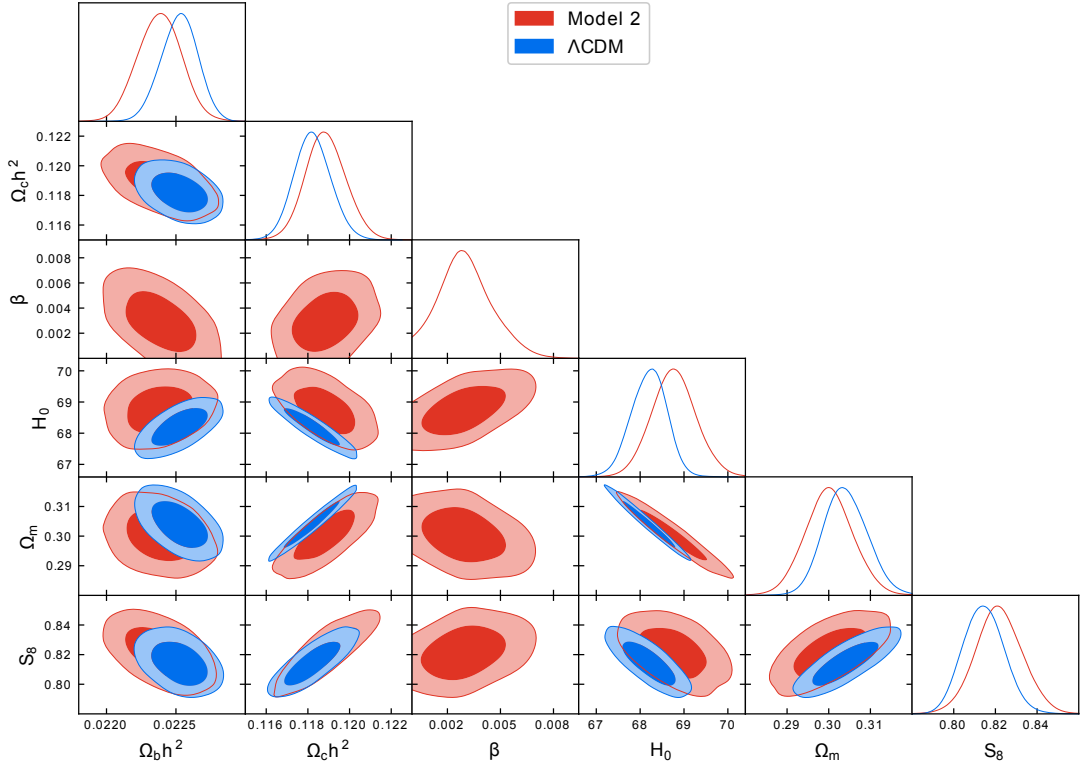}}
\caption{1D likelihoods and 2D contours for the parameters in 68\% and 95\% CLs for Model 2 (red) next to the $\Lambda$CDM constraints (blue).}
\label{figure:contour2}
\end{center}
\end{figure}

The 1D marginalized relative likelihood functions and 2D contours in 68\% and 95\% CLs for Model 2 are demonstrated in Figure \ref{figure:contour2}. In this case, the mean and best-fit values of $H_0$ are higher than the $\Lambda$CDM result. In the diagram of the 1D likelihood of $H_0$ it is also evident that the maximum likelihood of the Hubble parameter in this model is larger relative to the $\Lambda$CDM scenario. 2D contours for $H_0$ to all of the other parameters also show that the 68\% and 95\% CL marginalized joint regions are extended in Model 2 in comparison to $\Lambda$CDM. From these remarks, it seems that Model 2 can reduce the Hubble tension slightly in comparison with $\Lambda$CDM and also Model 1.

The best-fit value of the $\beta$ takes greater values in Model 2 compared to the two other particle production models. Thus, this scenario confirms the probability of particle production, more strongly. In this model the best-fit and mean values of $ \Omega_b h^2 $ and $ \Omega_c h^2 $ are smaller than the results of the $\Lambda$CDM model. The best-fit and mean values of the $ \Omega_m$ in this model are smaller than the ones in the $\Lambda$CDM model, and consequently the best-fit and mean values of the $\Omega_\Lambda$ in this model is obtained to be bigger than the $ \Lambda $CDM outcome. 

We see in Table \ref{table:parameters} that the best-fit of $S_8$ for Model 2 (0.8107) is slightly lower than that from the $\Lambda$CDM model (0.8162). So it seems that we can reduce the $S_8$ tension in Model 2 slightly relative to the $\Lambda$CDM scenario. But should be noted that the mean value from Model 2 (0.822) is higher than the $\Lambda$CDM result (0.8144). To explain this point, we note that the best-fit value from CosmoMC output is not the exact best-fit result of the model. Rather, it is the best model that has been hit by random walk so far, and it may be updated as the MCMC random walk continues. Although our assumption for the Gelman-Rubin convergence criterion ($ R-1<0.1$) is typically good enough for evaluating a marginalized distribution, it may not be good enough to give an accurate best-fit result. Thus, we cannot conclude definitely that Model 2 alleviates the $S_8$ parameter in comparison with the $\Lambda$CDM scenario, and caution needs to be taken for such a subtle comparison.

\begin{figure}[t]
\begin{center}
\scalebox{0.6}[0.6]{\includegraphics{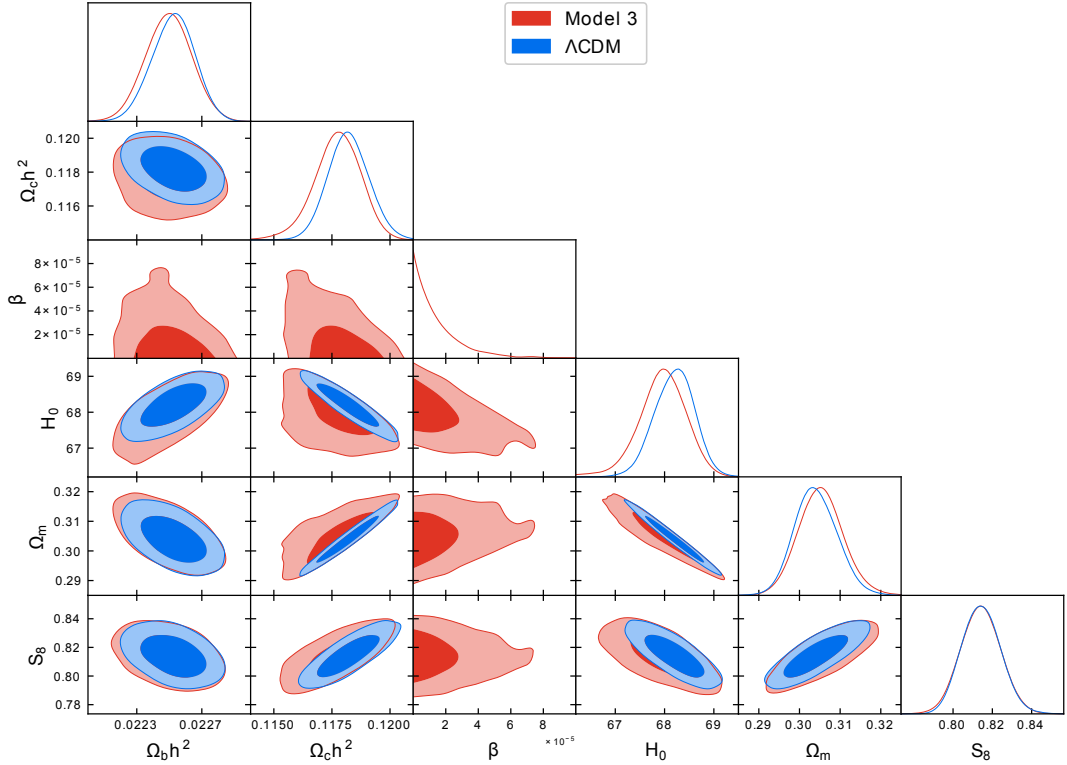}}
\caption{1D likelihoods and 2D contours for the parameters in 68\% and 95\% CLs for Model 3 (red) next to the $\Lambda$CDM constraints (blue).}
\label{figure:contour3}
\end{center}
\end{figure}

The 1D marginalized relative likelihood functions and 2D contours in 68\% and 95\% CLs for Model 3 are represented in Figure \ref{figure:contour3}. In this model, the joint analysis prepares strong constraints on all of the parameters too. Like Model 1, the best-fit of $\beta$ is very small in this case too. This small value however still confirms that the probability of particle production is consistent with the observations. The 68\% CL contour plot of this case is very similar to the $\Lambda$CDM joint regions. In fact, due to the small $\beta$, the behavior of this model is very similar to $\Lambda$CDM.

The best-fit value of $H_0$ in Model 3 is smaller than the $\Lambda$CDM result. Additionally, the 2D contours of this case in 68\% and 95\% CLs for $H_0$ are not substantially extended in comparison with the $\Lambda$CDM regions. Therefore, unlike Model 2, it seems that Model 3 cannot reduce the Hubble tension compared to the $\Lambda$CDM framework.

\begin{figure}
\resizebox{0.68\textwidth}{!}{
  \includegraphics{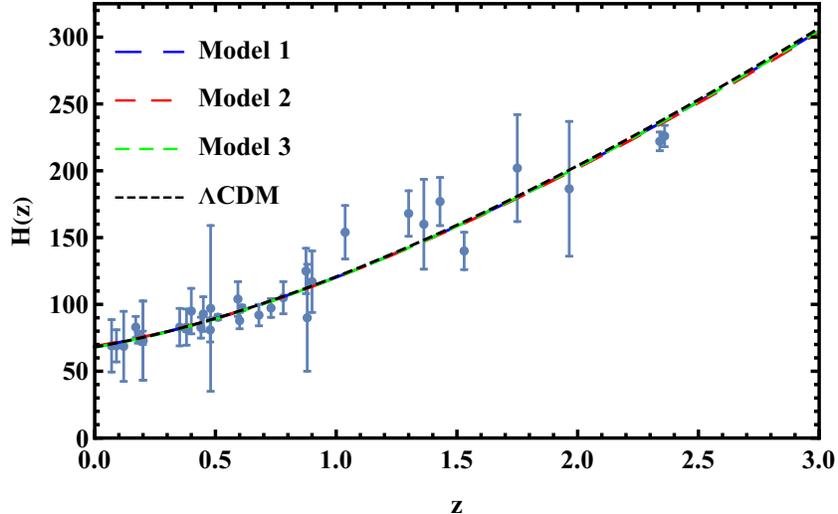}
}
\caption{Evolution of the Hubble parameter versus redshift in our scenario. The blue line, red dashed line, green dashed line, and black dashed line are corresponding to Model 1, Model 2, Model 3, and $\Lambda$CDM, respectively.}
\label{figure:Hubble}       
\end{figure}

\begin{figure}
\resizebox{0.68\textwidth}{!}{
  \includegraphics{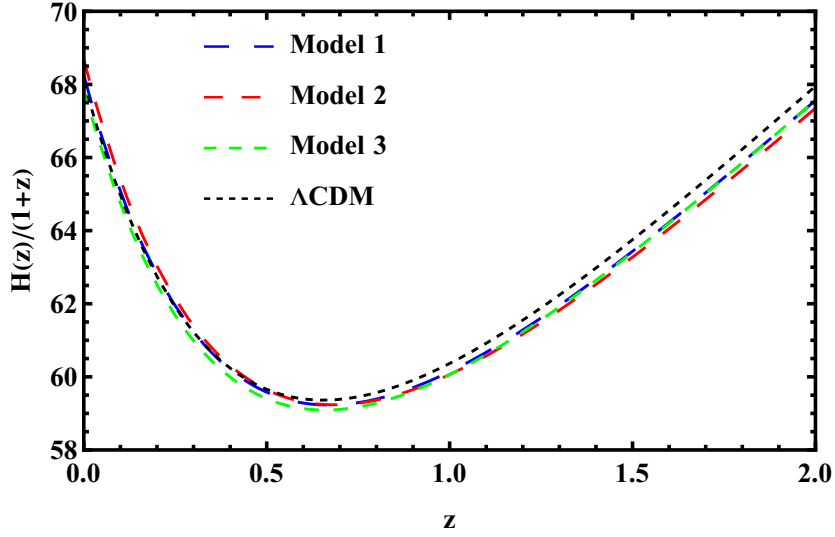}
}
\caption{Evolution of $ H(z)/(1+z) $ versus redshift in our scenario. The blue line, red dashed line, green dashed line, and black dashed line shows the results of Model 1, Model 2, Model 3, and $\Lambda$CDM, respectively.}
\label{figure:Hubble2}       
\end{figure}

In the following, we implement the best-fit values of the model parameters presented in Table \ref{table:parameters} to explore the behavior of the background cosmological quantities in our scenario during the Universe expansion. The most important background variable is the Hubble parameter which specifies the expansion rate of the Universe during its evolution. The diagram of this quantity in our scenario is demonstrated in Figure \ref{figure:Hubble}. In the figure, we have also compared the result of our models with that of the $\Lambda$CDM model as well as the cosmological data from the Hubble Space Telescope (HST). The data that we used in our work are presented in Table \ref{table:dataHST}. In the figure, we see that the results of the three cases of our scenario are very close to the one for the $\Lambda$CDM benchmark model. Since Model 1 and Model 3 are indistinguishable from $\Lambda$CDM in the background behavior, in Figure \ref{figure:Hubble2}, we plot $ H(z)/(1+z) $ to show the deviation of our models from $\Lambda$CDM more clearly. Although, the evolution of $H(z)$ in these models is very similar to the $\Lambda$CDM result, very small amounts of particle production rate cause $ \dot{a} $ evolves differently in these models from the $\Lambda$CDM model at some cosmological redshifts.

\begin{table}[!ht]
\caption{The observational data from HST that we used in our work.}
\centering
 \scalebox{0.9}{
\begin{tabular}{|l c c | l c c | l c c|}
\hline
Ref.  &  $\qquad z \qquad$ & $\qquad H (z) \qquad$&Refs.  & $\qquad z \qquad$ & $\qquad H(z) \qquad$ & Refs.  &  $\qquad z \qquad$ & $\qquad H(z) \qquad$  \\
\hline
\cite{Zhang2014}& 0.07 &$69.0 \pm 19.6$ & \cite{Moresco2016}&0.4783  &$ 80.9 \pm 9$&\cite{Simon2005}& 0.09 &$ 69 \pm 12 $ \\

\cite{Stern2010}&0.48&$97 \pm 62$& \cite{Zhang2014}& 0.12&$ 68.6   \pm 26.2$&\cite{Moresco2012}& 0.593  &$ 104 \pm 13$ \\

\cite{Simon2005} & 0.17 &$ 83 \pm 8 $&\cite{Moresco2012} & 0.68  &$ 92 \pm 8$&\cite{Moresco2012}& 0.179&$ 75 \pm 4$\\

\cite{Moresco2012}& 0.781    & $105 \pm 12$& \cite{Moresco2012}& 0.199  &$ 75 \pm 5$&\cite{Moresco2012} &0.875 &$125 \pm  17 $\\

\cite{Zhang2014} &0.20 & $72.9 \pm 29.6 $ &\cite{Stern2010}&0.88 & $90 \pm 40$&\cite{Simon2005}&0.27 & $77 \pm 14$\\

\cite{Simon2005}& 0.9 & $117 \pm 23$&\cite{Zhang2014}& 0.28 & $88.8 \pm 36.6 $&\cite{Moresco2012} &1.037      & $154 \pm 20$ \\

\cite{Moresco2012}& 0.352       & $83 \pm 14$&\cite{Moresco2016} &0.18 & $0.360 \pm 0.090 $ &\cite{Simon2005}&1.3         &$168 \pm 17$\\

\cite{Moresco2016}   &  0.3802       &$83 \pm 13.5 $ &\cite{Moresco2016}&1.363      &$ 160 \pm 33.6$&\cite{Moresco2016} & 0.57 &$0.417 \pm 0.045$\\

\cite{Simon2005} & 0.4      &$95 \pm 17 $&\cite{Simon2005}& 1.43       & $177 \pm 18$& \cite{Moresco2016}&0.4004      & $77 \pm 10.2$ \\

\cite{Simon2005} &1.53       &$ 140 \pm 14 $& \cite{Moresco2016}& 0.4247   &$87.1  \pm  11.2$&\cite{Moresco2016}& 1.75  &$202 \pm 40$\\

\cite{Moresco2016}& 0.44497   &$92.8  \pm 12.9 $&\cite{Moresco2016}& 0.38 &$0.477 \pm 0.051$&\cite{Moresco2015}& 1.965   &$186.5 \pm 50.4$\\
\hline 
\end{tabular}
}
\label{table:dataHST}
\end{table}

Although we have included the present Hubble parameter from the Riess et al. (2019) measurement \cite{Riess:2019cxk} in our CosmoMC analysis, however, it is useful here to evaluate the compatibility of our framework also with the local data from the HST measurements at different redshifts. The Hubble parameter is related to redshift independently of the theoretical model with the following relation
\begin{equation}
H(z)=-\frac{1}{1+z}\frac{{\rm d}z}{{\rm d}t}.
\end{equation}
So if ${\rm d}z/{\rm d}t$ is known, $H(z)$ can be determined directly \cite{Jimenez}. The best-fit values of the model parameters from HST can be determined by minimizing \cite{Samushia}
\begin{equation}
\chi_{\rm HST}^{2}=\sum_{\rm i}\frac{\left[H_{\rm obs}(z_{\rm
i})-H_{\rm th}(z_{\rm i},\textbf{q})\right]^{2}}{\sigma_{\rm
i}^{2}}.
\end{equation}
The observational data for the Hubble parameter in the redshift interval of $0.07 \leq z \leq 1.965$ are listed in Table \ref{table:dataHST}. We use these data points to estimate the value of $\chi_{\rm HST}^{2}$ for the different cases of our scenario as reported in Table \ref{table:chi2HST}. From the table, we deduce that the $\Lambda$CDM fits the HST data better than our three particle production models. Among the particle production models, Model 1 provides a better fit with the data in comparison with the other two cases.

\begin{table}
\caption{Values of $\chi_{\mathrm{HST}}^{2}$ for the particle production models in comparison with the $\Lambda$CDM result. In the table, we also report the values of $\Delta\chi_{\mathrm{HST}}^{2}\equiv\chi_{\mathrm{HST(Model)}}^{2}-\chi_{\mathrm{HST(\Lambda CDM)}}^{2}$.}
\begin{tabular}{|c|c|c|c|c|}
\hline 
$\qquad$ $\qquad$ & $\qquad$ Model 1 $\qquad$ & $\qquad$ Model 2 $\qquad$ & $\qquad$ Model 3 $\qquad$ & $\qquad$ $\Lambda$CDM $\qquad$ \\ 
\hline 
$\chi_{\mathrm{HST}}^{2}$ & $14.4605$ & $14.4732$ & $14.5611$ & $14.3731$ \\
$\Delta\chi_{\mathrm{HST}}^{2}$ & $0.0874$ & $0.1001$ & $0.188$ & $0.0$ \\
\hline 
\end{tabular}
\label{table:chi2HST}
\end{table}

\begin{figure}
\resizebox{0.65\textwidth}{!}{%
  \includegraphics{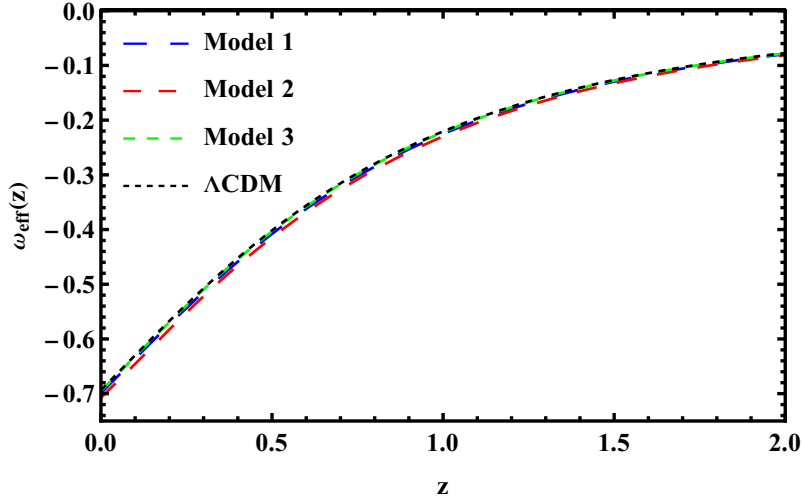}
}
\caption{The effective EoS parameter in our particle production cosmological scenario compared to the standard $\Lambda$CDM cosmology, using the best-fit value of model parameters presented in Table \ref{table:parameters}.}
\label{figure:EoS}      
\end{figure}

By using the best-fit values of the parameters in Table \ref{table:parameters}, we then plot the variation of the effective equation of state parameter $ \omega_\mathrm{eff}(z) $ in against of cosmological redshift. The diagram of this quantity is drawn in Figure \ref{figure:EoS} for the particle production models together with the $\Lambda$CDM plot. By regarding the best value of the parameters provided in Table \ref{table:parameters}, we find the present value of the EoS as $\omega_{\mathrm{eff}} = -0.699$ for Model 1, $\omega_{\mathrm{eff}} = -0.707$ for Model 2, and $\omega_{\mathrm{eff}} = -0.695$ for Model 3. These values are close to EoS of $\Lambda$CDM with $\omega_{\mathrm{eff}} = -0.694$. The result of Model 3 is very close to the one of $\Lambda$CDM because the best-fit value of the $ \beta $ parameter is very tiny in this case. It is worthwhile to remind here that as the parameter $\beta$ takes smaller values in our setting, the behavior of that model tends further to the treatment of the standard $\Lambda$CDM model without particle production. Model 1 and Model 2, however, offer more negative values for $\omega _0$ relative to $\Lambda$CDM.

\begin{figure}
\resizebox{0.65\textwidth}{!}{%
  \includegraphics{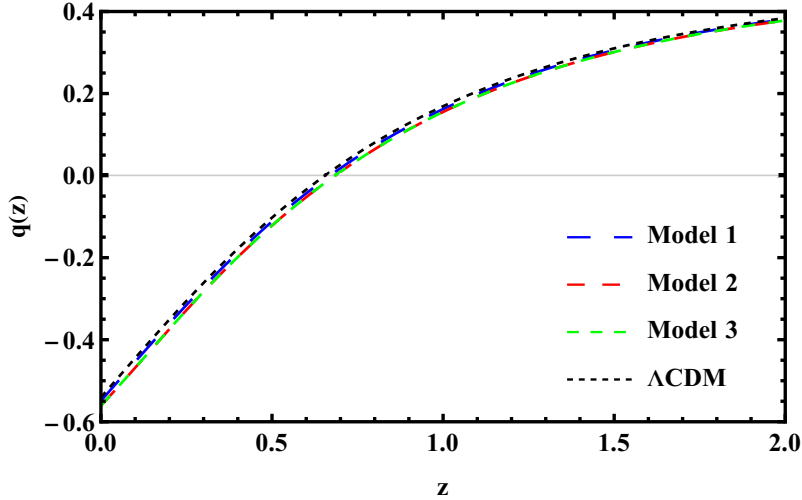}
}
\caption{Deceleration parameter of our particle production models and the $\Lambda$CDM cosmology, using the best-fit value of model parameters presented in Table \ref{table:parameters}.}
\label{figure:q}       
\end{figure}

In Figure \ref{figure:q}, we present the redshift evolution of the deceleration parameter, $q(z)$, for the three models including particle production accompanied by the plot of the $\Lambda$CDM cosmology. The figure indicates that the Universe enters to the accelerated phase of expansion at the transition redshift $z_t = 0.669$ for Model 1, $z_t = 0.684$ for Model 2, $z_t = 0.684$ for Model 3, and $z_t = 0.655$ for $\Lambda$CDM. Therefore, the three investigated models involving particle production begin the accelerated phase of expansion earlier than $\Lambda$CDM. The present value of the deceleration parameter is obtained as $ q_0= -0.55$ for Model 1, $ q_0= -0.56$ for Model 2, $ q_0= -0.56$ for Model 3, and $ q_0= -0.54$ for $\Lambda$CDM. So, the deceleration parameter in three investigated models is more negative than the $\Lambda$CDM deceleration parameter.

\section{Growth factor}
\label{section:growth_factor}
 
\begin{figure}
\resizebox{0.65\textwidth}{!}{%
  \includegraphics{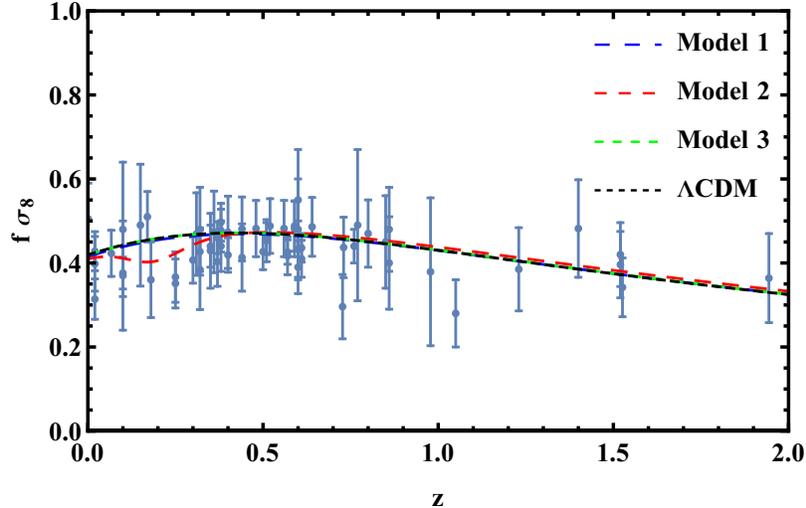}
}
\caption{Evolution of the growth factor with redshift in Model 1, Model 2, Model 3, and  $\Lambda$CDM, in comparison with the observational data of Table \ref{table:datafsigma8}.}
\label{fsigma}       
\end{figure}

In order to investigate the evolution of density perturbation in the linear regime, we solve numerically the quadratic differential equation of $\delta$ for each case of our model by using the best-fit values of its parameters reported in Table \ref{table:parameters}. Since, the space-time perturbations are assumed to be adiabatic during the Universe evolution in our scenario, then we take $c_\mathrm{eff}^2 \approx c_s^2$ \cite{Rezazadeh:2020zrd}. With this assumption, one can solve the coupled differential equations \eqref{ddotdeltab} and \eqref{ddotdeltac}, numerically. Equation \eqref{ddotdeltac} describes the evolution of the cold dark matter growth factor, and in general, its solutions are scale-dependent, as it is evident from the equation. Since the equation of state in this case is non-zero, it is useful to introduce density contrast as Fourier modes $ \delta=\int \delta_k  e^{i\vec{k}.\vec{r}}d^3 k $ to solve the differential equation. Therefore, $ \nabla^2 \delta = - k^2 \delta $, where $k$ is comoving wavenumber, relates each Fourier mode to a wavenumber. Using different amounts of $k$, we checked that the results are approximately $k$-independent. In our numerical computations, we fixed the comoving wavenumber to $k=0.01\,h\,\mathrm{Mpc}^{-1}$ which is deep inside the Hubble horizon during the interested cosmological redshifts, and also it is related to the structures which are in the linear regime of perturbations during the interested cosmological eras. Using the solutions of Eqs. \eqref{ddotdeltab} and \eqref{ddotdeltac}, we can evaluate the following quantities
\begin{align}
 \label{fz}
 f(z) &=\frac{d\ln\delta_{m}}{d\ln a},
 \\
 \label{sigma8z} 
 \sigma_{8}(z) &=\frac{\delta_{m}(z)}{\delta_{m}(z=0)}\sigma_{8}(z=0).
\end{align}
By using these quantities, we can calculate the growth factor $f(z)\sigma_{8}(z)$ whose diagram is displayed in Fig. \ref{fsigma} in terms of redshift for all the investigated models. In the figure, we also displayed the observational data summarized in Table \ref{table:datafsigma8}. Although all the three models including particle production behave like $\Lambda$CDM at high redshifts, the deviation from $\Lambda$CDM becomes more pronounced at low redshifts. Model 1 and Model 3 remain close to $\Lambda$CDM also at the low redshifts, whereas Model 2 reveals significant deviation from $\Lambda$CDM at these redshifts. Strictly speaking, the deviation at low redshifts leads to a better consistency with the LSS observations for Model 2 in comparison with the $\Lambda$CDM scenario. We see in the figure that Model 2 covers a few data points that $\Lambda$CDM is not able to support. Using the best-fit value of the parameters provided in Table \ref{table:parameters}, we calculate the $\chi^2_{f \sigma_8}$ for each case of three frameworks and summarize the findings in Table \ref{table:chi2fsigma8}. It is noteworthy that all three models give lower values for $\chi^2_{f \sigma_8}$ compared to the $\Lambda$CDM model. Model 2 gives the minimum value of $\chi^2_{f \sigma_8}$ between the investigated models, and its better consistency results from the late-time effects of the particle production in this model on the sound speed of perturbations. Since, this model performs better than the other models at both the background and perturbations levels, we refer it as the best case of our particle production scenario. It should be noted that in this case, the rate of the particle production is assumed to become more pronounced after a special fixed scale factor as it can be deduced from its functional form in Eq. \eqref{Model2}. It will be more general if the critical scale factor of the particle production is allowed to be a free parameter that should be determined by the MCMC analysis, and we leave the study of this possibility for future investigations.

\begin{table}[!ht]
\caption{The observational data for the growth factor $f \sigma_8 (z) $ that we used in this work.}
\centering
 \scalebox{0.9}{
\begin{tabular}{|c c c | c c c | c c c|}
\hline
Ref.  &  $\qquad z \qquad$ & $\qquad f \sigma_8 (z) \qquad$ & Ref.  & $\qquad z \qquad$ & $\qquad f \sigma_8 (z) \qquad$ & Ref.  & $\qquad z \qquad$ & $\qquad f \sigma_8 (z) \qquad$  \\
\hline
\cite{Song}& 0.35 &$0.440 \pm 0.050$ & \cite{Song}& 0.77 &$0.490 \pm 0.18$&\cite{Song}& 0.17&$ 0.510 \pm 0.060 $ \\

\cite{Davis, Hudson}&0.02 &$0.314 \pm 0.048 $&\cite{Hudson, Turnbull}& 0.02&$ 0.398 \pm 0.065 $&\cite{Samushia2}& 0.25&$ 0.3512 \pm 0.0583$ \\

\cite{Samushia2} & 0.37&$ 0.4602 \pm 0.0378 $&\cite{Samushia2} & 0.25&$ 0.3665 \pm 0.0601$&\cite{Samushia2}& 0.37&$ 0.4031 \pm 0.0586$\\

\cite{Larson}& 0.44& $0.413 \pm 0.080$& \cite{Larson}& 0.60&$ 0.390 \pm 0.063$&\cite{Larson} &0.73 &$0.437 \pm 0.072 $\\

\cite{Beutler2} &0.067& $0.423 \pm 0.055 $ &\cite{Tojeiro}&0.30& $0.407 \pm 0.055$&\cite{Tojeiro}&0.40& $0.419 \pm 0.041$\\

\cite{Tojeiro}& 0.50& $0.427 \pm 0.043$&\cite{Tojeiro}& 0.60& $0.433 \pm 0.067 $&\cite{Torre} &0.80& $0.470 \pm 0.080$ \\

\cite{Chuang}& 0.35& $0.429 \pm 0.089$&\cite{Blake} &0.18 & $0.360 \pm 0.090 $ &\cite{Blake}&0.38 &$0.440 \pm 0.060$\\

\cite{Sanchez}   &  0.32 &$0.384 \pm 0.095 $ &\cite{Sanchez}& 0.32&$ 0.48 \pm 0.10$&\cite{Sanchez} & 0.57 &$0.417 \pm 0.045$\\

\cite{Howlett} & 0.15 &$0.490 \pm 0.145 $&\cite{Feix}& 0.10& $0.370 \pm 0.130$& \cite{Okumura}&1.40& $0.482 \pm 0.116$ \\

\cite{ Chuang2013} &0.59&$ 0.488 \pm 0.060 $& \cite{Alam}& 0.38 &$0.497 \pm 0.045$&\cite{Alam}& 0.51 &$0.458 \pm 0.038$\\

\cite{Alam}& 0.61 &$0.436 \pm 0.034 $&\cite{Beutler2017}& 0.38 &$0.477 \pm 0.051$&\cite{Beutler2017}& 0.51 &$0.453 \pm 0.050$\\

\cite{Beutler2017}& 0.61 &$0.410 \pm 0.044$&\cite{Wilson}&0.76 &$0.440 \pm 0.040 $&\cite{Wilson} & 1.05 &$0.280 \pm 0.080 $\\

\cite{Marin}  & 0.32 &$0.427 \pm 0.056$&\cite{Marin}& 0.57 &$0.426 \pm 0.029$&\cite{Hawken}&0.727 &$0.296 \pm 0.0765$\\

\cite{Huterer} & 0.02&$ 0.428 \pm 0.0465 $&\cite{Torre2017}&0.6 &$0.48 \pm 0.12$&\cite{Torre2017} &0.86 &$0.48 \pm 0.10 $\\

\cite{Pezzotta} &0.60& $0.550 \pm 0.120 $& \cite{Pezzotta}& 0.86&$0.400 \pm 0.110 $&\cite{Feix2017}  & 0.1 &$0.48 \pm 0.16$ \\

\cite{Howlett2017} &0.001&$0.505 \pm 0.085$&\cite{Mohammad} & 0.85 &$0.45 \pm 0.11 $&\cite{Alam2015}& 0.31 &$0.469 \pm 0.098$ \\

\cite{Alam2015}& 0.36 &$0.474 \pm 0.097 $&\cite{Alam2015} & 0.40 &$0.473 \pm 0.086 $&\cite{Alam2015}& 0.44 &$0.481 \pm 0.076 $\\

\cite{Alam2015} & 0.48 &$0.482 \pm 0.067 $&\cite{Alam2015} & 0.52 &$0.488 \pm 0.065$&\cite{Alam2015}& 0.56 &$0.482 \pm 0.067 $\\

\cite{Alam2015} & 0.59 &$0.481 \pm 0.066 $&\cite{Alam2015} & 0.64 &$0.486 \pm 0.070$&\cite{Shi2018} & 0.1& $0.376 \pm 0.038 $\\

\cite{Marin2018} & 1.52&$ 0.420 \pm 0.076 $&\cite{Hou2018}  &1.52&$ 0.396 \pm 0.079$&\cite{Zhao} & 0.978 &$0.379 \pm 0.176 $\\

\cite{Zhao} &1.23& $0.385 \pm 0.099$&\cite{Zhao} &1.526 &$0.342 \pm 0.070 $&\cite{Zhao} &1.944& $0.364 \pm 0.106$\\
\hline 
\end{tabular} 
}
\label{table:datafsigma8}
\end{table}

\begin{table}[!ht]
\caption{The minimum of $\chi^2_{f \sigma_8}$ for the three models involving particle production together with the result of the $\Lambda$CDM scenario without particle production. In the table, we also report the values of $\Delta\chi_{f\sigma_{8}}^{2}\equiv\chi_{f\sigma_{8}\mathrm{(Model)}}^{2}-\chi_{f\sigma_{8}\mathrm{(\Lambda CDM)}}^{2}$.}
\centering
\begin{tabular}{|c|c|c|c|c|}
\hline
$\qquad$ $\qquad$  & $\qquad$ Model 1 $\qquad$ & $\qquad$ Model 2 $\qquad$ & $\qquad$ Model 3 $\qquad$ & $\qquad$ $\Lambda$CDM $\qquad$ \\ 
\hline 
$\chi^2_{f \sigma _8}$ & $42.9220$ & $39.8524$ & $45.3846$ & $45.2861$ \\
$\Delta\chi_{f\sigma_{8}}^{2}$ & $-2.3641$ & $-5.4337$ & $0.0985$ & $0.0$ \\
\hline
\end{tabular} 
\label{table:chi2fsigma8}
\end{table}


\section{Conclusion}
\label{section:conclusions}

In this work, we studied a cosmological setup that involves the dark matter self-interactions during the evolution of the Universe. By assuming the Universe as an open thermodynamic system and applying the concepts of non-equilibrium thermodynamics, we studied the cosmological implications of the process of gravitational particle production. In our research, we considered three profiles for the particle production rate which has well-based theoretical and phenomenological motivations. We studied the large-scale structure formation in our cosmological setup and specially extracted the equations governing the dark matter overdensities in the linear regime of perturbations. In the limit of vanishing particle prediction rate, our results reduce to the well-known equations. In the next step, we applied the cosmological data from the Planck 2018 measurements for the anisotropies observed in the temperature and polarization spectra of CMB radiation \cite{Planck:2018vyg, Planck:2019nip, Planck:2018lbu}, the Pantheon SNI survey \cite{Pan-STARRS1:2017jku}, the BAO observations \cite{BOSS:2016wmc, Ross:2014qpa, Beutler:2012px}, the Riess et al. (2019) constraint for the Hubble constant \cite{Riess:2019cxk}, to put observational constraints on the parameters of our model. For this purpose, we run the CosmoMC code \cite{Lewis2000, Lewis2002} which is based on the MCMC numerical method. For the statistical analysis of the CosmoMC chains, we utilized the GetDist computational package \cite{Lewis:2019xzd}. The numerical results of CosmoMC are presented in Table \ref{table:parameters} and Table \ref{table:chi2}. Also, the two-dimensional contour plots obtained from this code are demonstrated in Figs. \ref{figure:contour1}-\ref{figure:contour3}. The code gives the minimum value of $\chi_{{\rm tot}}^{2}$ as $3838.99$ for Mode 1, $3834.40$ for Model 2, $3837.33$ for Model 3, and $3838.00$ for $\Lambda$CDM. Therefore, Model 2 and Model 3 in our cosmological particle production scenario fit the observational data better than the $\Lambda$CDM model, while Model 1 fails to provide a better fit with observations than the standard cosmological scenario.

In the three models introduced in this paper, we obtained the particle production rate using cosmological constraints, $\Gamma/3 H_0= 6.0\times 10^{-5}$ for Model 1, $\Gamma/3 H_0= 0.01719$ for Model 2 and $\Gamma/3 H_0= 1.0 \times10^{-6}$ for Model 3. Therefore, the possibility of the particle production is approved as consistent with recent cosmological observations.

The contribution of the dark energy component in all three models involving particle production obtains greater amounts than the $\Lambda$CDM contribution. Also, the models involving particle production result in greater optical depth than $\Lambda$CDM without particle production.

We found the 68\% CL constraint for the Hubble constant in our setup as $H_{0}=67.96_{-0.40}^{+0.50}\,{\rm km\,s^{-1}\,Mpc^{-1}}$, $H_{0}=68.79\pm0.59\,{\rm km\,s^{-1}\,Mpc^{-1}}$, and $H_{0}=67.93_{-0.41}^{+0.53}\,{\rm km\,s^{-1}\,Mpc^{-1}}$ for Model 1, Model 2, and Model 3, respectively. Thus, Model 2, in contrast to Model 1 and Model 3, provides a larger value for the Hubble constant compared to the $\Lambda$CDM model giving $H_{0}=68.20_{-0.38}^{+0.42}\,{\rm km\,s^{-1}\,Mpc^{-1}}$, and therefore, we can reduce the Hubble tension slightly in our scenario.

We checked the consistency of the particle production cosmology with the HST data which are related to the local surveys of the Hubble parameter in the redshift interval $0.07 \leq z \leq 1.965$. Our analysis implies that although Model 2 and Model 3 in our scenario provide a better fit with the CMB, SNI, BAO, and Riess et al. (2019) data, the standard $\Lambda$CDM cosmology fits the HST data better than the three cases of particle production in our setting. From the evolution of $H(z)/(1+z)$ versus redshift, we concluded that the deviation in the background dynamics relative to $\Lambda$CDM is more clear at the high redshifts, and in these redshifts, the particle production models result in smaller values for the Hubble parameter. 

The diagram of the deceleration parameter $q(z)$ indicates that our particle production scenario arrives at the accelerated phase of the Universe earlier than $\Lambda$CDM. Also, the present values of EoS in all three investigated models involving particle production get more negative amounts than the $\Lambda$CDM result. $\omega_{\mathrm{eff}}$ in Model 3 is very close to the $\Lambda$CDM result, due to a tiny value of $\beta$.

Then, we focused on the implications of our scenario in light of the linear perturbations quantities. We solved the equations of the density contrasts numerically, and used the solutions to estimate the growth factor $f\sigma_{8}(z)$. We assessed our findings in light of the cosmological data from the LSS measurements. The value of $\chi_{f\sigma_{8}}^{2}$ for the utilized data sample is acquired as $42.92$, $39.85$, $45.38$, and $45.29$ for Model 1, Model 2, Model 3, and $\Lambda$CDM, respectively. Therefore, particle production models 1 and 2 yield a better fit to the LSS data in comparison with the $\Lambda$CDM model. The best result belongs to Model 2 which also provides a better fit with the CMB, SNI, BAO, and Riess et al. (2019) data at the level of background dynamics. The better compatibility of Model 2 with the observations compared to $\Lambda$CDM at the background level is to some extent due to the one addition degree freedom ($\beta$), but at the perturbation level, its better performance arises primarily from the impact of the cosmological particle production on the sound speed of inhomogeneities at low redshifts. Since Model 2 improves the fitting with the background and perturbation data, we refer to this model as the best case in our particle production scenario that deserves more studies in future investigations. Specifically, it should be noted that in the functional form that we regarded for the production rate of this case in the present work, we restricted the particle production to be more efficient after a fixed special scale factor, as it can be deduced from Eq. \eqref{Model2}. As an important extension, we can consider a generalized form for the production rate of this case such that the particle production scale factor to be a free parameter and be determined by the MCMC analysis. Such a generalization may provide even a better fit with the observational data for the model. In this way, we may also be able to reduce the $H_0$ and $S_8$ tension more efficiently. We leave the study of this possibility for future works. Furthermore, it will much useful to study the non-linear regime of perturbations in our particle production scenario and investigate their cosmological implications. We leave the study of this issue for future investigations too.




\bibliographystyle{aip}


\end{document}